\documentclass[12pt,a4j]{article}
\usepackage{amsmath,amsthm,amssymb,bm,fancybox,multirow,hhline,psfrag,comment}
\usepackage{url}
\usepackage{color}
\usepackage[dvipdfm]{graphicx}
\usepackage{natbib}

\def\hsymbu#1{\smash{\lower1.7ex\hbox{\huge$#1$}}}

\usepackage{indentfirst}
\usepackage{algorithm,algorithmic}

\newtheorem{Example}{Example}[section]
\bibpunct{(}{)}{;}{a}{}{,}

\def\E{{E}}

\def\l{{l}}

\makeatletter
\def\eqnarray{\stepcounter {equation}\let \@currentlabel =\theequation
\global \@eqnswtrue
\global \@eqcnt \z@ \tabskip \@centering \let \\=\@eqncr
$$\halign to \displaywidth \bgroup \@eqnsel \hskip \@centering
$\displaystyle \tabskip \z@ {##}$&\global \@eqcnt \@ne \hfil
${\mbox{}##\mbox{}}$\hfil &\global \@eqcnt \tw@
$\displaystyle \tabskip \z@ {##}$\hfil \tabskip \@centering
&\llap {##}\tabskip \z@ \cr}
\makeatother

\begin{document}

\theoremstyle{plain}
\newtheorem{lemma}{Lemma}[section]
\theoremstyle{remark}
\newtheorem{remark}{Remark}[section]
\theoremstyle{example}
\newtheorem{example}{Example}[section]

{
\begin{center}
\textbf{\Large Estimation of oblique structure via penalized likelihood factor analysis}
\end{center}
\begin{center}
\large {Kei Hirose and Michio Yamamoto 
}
\end{center}

\begin{center}
{\it {\small
Division of Mathematical Science, Graduate School of Engineering Science, Osaka University,\\
1-3, Machikaneyama-cho, Toyonaka, Osaka, 560-8531, Japan \\



}}

\vspace{1.5mm}

{\it {\small E-mail: hirose@sigmath.es.osaka-u.ac.jp, \ myamamoto@sigmath.es.osaka-u.ac.jp.\\
}}

\end{center}

\begin{abstract}
We consider the problem of sparse estimation via a lasso-type penalized likelihood procedure in a factor analysis model.  Typically, the model estimation is done under the assumption that the common factors are orthogonal (uncorrelated).  However, the lasso-type penalization method based on the orthogonal model can often estimate a completely different model from that with the true factor structure when the common factors are correlated.  In order to overcome this problem, we propose to incorporate a factor correlation into the model, and estimate the factor correlation along with parameters included in the orthogonal model by maximum penalized likelihood procedure.  An entire solution path is computed by the EM algorithm with coordinate descent, which permits the application to a wide variety of convex and nonconvex penalties.  The proposed method can provide sufficiently sparse solutions, and be applied to the data where the number of variables is larger than the number of observations.  Monte Carlo simulations are conducted to investigate the effectiveness of our modeling strategies.  The results show that the lasso-type penalization based on the orthogonal model cannot often approximate the true factor structure, whereas our approach performs well in various situations.  The usefulness of the proposed procedure is also illustrated through the analysis of real data.
 \end{abstract}

\section{Introduction}

Factor analysis provides a practical tool for exploring the covariance structure among a set of observed random variables by construction of a smaller number of random variables called common factors. 
In exploratory factor analysis, a traditional estimation procedure in use is the following two-step approach: the model is estimated by the maximum likelihood method under the assumption that the common factors are uncorrelated (orthogonal),  and then rotation techniques, such as the varimax method \citep{kaiser1958varimax} and the promax method \citep{hendrickson1964promax}, are utilized to find sparse factor loadings.   
However, it is well known that the maximum likelihood method often yields unstable estimates because of overparametrization (e.g., \citealp{akaike1987factor}).  In particular, the commonly-used algorithms for maximum likelihood factor analysis (e.g., \citealp{joreskog1967some}; \citealp{jennrich1969newton}; \citealp{clarke1970rapidly}; \citealp{lawley1971factor}) cannot often be applied when the number of variables is larger than the number of observations.  Furthermore,  the rotation techniques cannot often produce a sufficiently sparse solution.  In order to overcome these difficulties, we apply a penalized likelihood procedure that produces the sparse solutions, such as the lasso \citep{Tibshirani:1996}.  
The lasso-type penalized likelihood factor analysis has been recently studied by several researchers. \citet{Ningetal:2011} and \citet{Choietal:2011} applied the weighted lasso to obtain sparse factor loadings, and  numerically demonstrated that the penalization method often outperformed the rotation technique with maximum likelihood procedure.  \citet{hirose2012sparse} showed that the penalization method is a generalization of the rotation technique with maximum likelihood method, and applied the nonconvex penalties such as minimax concave penalty (MC+, \citealp{Zhang:2010}) and smoothly clipped absolute deviation (SCAD, \citealp{FanLi:2001}) to achieve sparser solutions than the lasso.  

In these studies, the common factors are assumed to be uncorrelated (orthogonal) as is the case with  the maximum likelihood exploratory factor analysis.   In some cases, however, analysts may prefer to relax the requirement that the common factors are orthogonal  (e.g., \citealp{mulaik1972foundations}).    Moreover, we found that the lasso-type penalization technique based on the orthogonal model can often estimate a completely different model from that with the true factor structure when the common factors are correlated (oblique).  Empirically,  the estimated factor loadings in the first column often become dense (i.e., all elements are non-zero), even if the first column of true loading matrix is sparse.  

In order to handle this fundamental problem,  we propose to incorporate a factor correlation into the model, and estimate the factor correlation  along with  parameters included in the orthogonal model by  maximum penalized likelihood procedure.  A pathwise algorithm via the EM algorithm \citep{rubin1982algorithms} with coordinate descent for nonconvex penalties \citep{Mazumderetal:2009} is introduced according to the basic idea given by \citet{hirose2012sparse}.  Our algorithm produces the entire solution path for a wide variety of convex and nonconvex penalties including the lasso, SCAD, and MC+ family.      Furthermore, the proposed methodology can provide sparser solutions than the rotation technique with maximum likelihood method, and be applied to the data where the number of variables is larger than the number of observations.   

The remainder of this paper is organized as follows: Section 2 shows that the lasso-type penalized likelihood factor analysis   based on the orthogonal model cannot often approximate the oblique structure.   In Section 3,  we introduce a penalized factor analysis via the oblique model, and provide a computational algorithm based on the EM algorithm and coordinate descent to obtain the entire solution path.   Section 4 presents numerical results for both artificial and real datasets. Some concluding remarks are given in Section 5.


\section{Penalized likelihood factor analysis  based on the orthogonal model may not approximate the oblique structure} \label{sec:orthogonal does not work}
\subsection{Model and Estimation}
We briefly describe  a lasso-type penalized likelihood factor analysis based on the orthogonal model (\citealp{Choietal:2011,Ningetal:2011,hirose2012sparse}).       
Suppose that  $\bm{X}=(X_1,\dots,X_p)^T$ is a $p$-dimensional observable random vector with mean vector $\bm{\mu} $ and variance-covariance matrix $\mathbf{\Sigma}$. 
The factor analysis model (e.g., \citealp{mulaik1972foundations}) is
\begin{equation*}
\bm{X} =\bm{\mu} + \mathbf{\Lambda} \bm{F}+\bm{\varepsilon}  , \label{model1}
\end{equation*}
where $\mathbf{\Lambda} =(\lambda_{ij})$ is a $p \times m$  matrix of factor loadings, and $\bm{F} = (F_1,\cdots,F_m)^T$ and $\bm{\varepsilon}  = (\varepsilon_1,\cdots, \varepsilon_p)^T$ are unobservable random vectors. The elements of $\bm{F}$  and $\bm{\varepsilon}$  are called common factors and unique factors, respectively. It is assumed that the common factors $\bm{F}$ and the unique factors $\bm{\varepsilon}$ are multivariate-normally distributed with  $\E(\bm{F} )=\mathbf{0}$,  $\E(\bm{\varepsilon} )=\mathbf{0}$,  $\E(\bm{F}\bm{F}^T)=\mathbf{I}_m$, $\E(\bm{\varepsilon} \bm{\varepsilon} ^T)=\mathbf{\Psi} $, and are independent (i.e., $\E(\bm{F} \bm{\varepsilon} ^T)=\mathbf{O}$). Here $\mathbf{I}_m$ is the $m \times m$ identity matrix, and $\mathbf{\Psi} $ is a $p \times p$ diagonal matrix with the $i$-th diagonal element $\psi_i$, which is called unique variance.     Under these assumptions, the observable random vector $\bm{X}$ is multivariate-normally distributed with variance-covariance matrix $\mathbf{\Lambda} \mathbf{\Lambda}^T+\mathbf{\Psi}$.  

Let $\bm{x}_1,\cdots,\bm{x}_N$  be a random sample of $N$ observations from the $p$-dimensional normal population $N_p(\bm{\mu} ,  \mathbf{\Lambda} \mathbf{\Lambda}^T+\mathbf{\Psi}) $.  The estimates of factor loadings and unique variances, say,  $\hat{\mathbf{\Lambda}}_{\rm ort}$ and $\hat{\mathbf{\Psi}}_{\rm ort}$ (``ort" is an abbreviation for orthogonal),  are obtained by maximizing the penalized log-likelihood function 
\begin{equation*}
(\hat{\mathbf{\Lambda}}_{\rm ort},  \hat{\mathbf{\Psi}}_{\rm ort}) = \mathrm{arg}\max_{\mathbf{\Lambda},\mathbf{\Psi}}\ell_{\rho}^{\rm ort}(\mathbf{\Lambda},\mathbf{\Psi}),\label{problem_rotation_pmle3_adj_ort}
\end{equation*}
where $\ell_{\rho}^{\rm ort}(\mathbf{\Lambda},\mathbf{\Psi})$ is the penalized log-likelihood function
\begin{equation*}
\ell_{\rho}^{\rm ort}(\mathbf{\Lambda},\mathbf{\Psi})  =  \ell^{\rm ort}(\mathbf{\Lambda},\mathbf{\Psi})  -N   \sum_{i=1}^p\sum_{j=1}^m \rho P(|\lambda_{ij}|).  \label{problem_rotation_pmle3_adj0_ort}
\end{equation*}
 Here $\ell^{\rm ort}(\mathbf{\Lambda},\mathbf{\Psi})$ is the log-likelihood function
\begin{eqnarray*}
\ell^{\rm ort}(\mathbf{\Lambda}} ,{\mathbf{\Psi})= -\frac{N}{2} \left[ p\log(2\pi)+\log |\mathbf{\Lambda} \mathbf{\Lambda}^T+\mathbf{\Psi}| + \mathrm{tr} \{ (\mathbf{\Lambda} \mathbf{\Lambda}^T+\mathbf{\Psi})^{-1} \mathbf{S} \} \right], \label{taisuuyuudo}
\end{eqnarray*}
$P(\cdot)$ is a penalty function, and $\rho$ is a regularization parameter.  The matrix $\mathbf{S}=(s_{ij})$ is the sample variance-covariance matrix. 
 
The lasso-type penalty function $P(\cdot)$ produces sparse solutions for some $\rho$, i.e., some of the factor loadings can be estimated by exactly zero. The lasso is continuous and fast, but biased and then estimates an overly dense model \citep{Zou:2006,zhao2007model,Zhang:2010}. Typically, a nonconcave penalization procedure such as MC+ \citep{Zhang:2010} and SCAD \citep{FanLi:2001} can achieve sparser models than the lasso.  For example, the MC+ \citep{Zhang:2010} is given by   
\begin{eqnarray*}
\rho P(|\theta|;\rho;\gamma)&=&\rho \int_0^{|\theta|}\left(1-\frac{x}{\rho\gamma}\right)_+dx\\
&=&\rho \left(|\theta|-\frac{\theta^2}{2\rho\gamma}\right) I(|\theta| < \rho\gamma) + \frac{\rho^2\gamma}{2}I(|\theta| \ge \rho\gamma).
\end{eqnarray*}
For each value of $\rho>0$, $\gamma \rightarrow \infty$ yields soft threshold operator (i.e., lasso penalty) and $\gamma \rightarrow 1+$ produces hard threshold operator.

\subsection{Problem of the lasso via orthogonal model} \label{sec:Problem of orthogonal model}
The lasso-type penalization based on the orthogonal model can perform well when the true common factors are uncorrelated.  In practical situations, however, the true common factors may often be correlated: $E[\bm{F}\bm{F}^T] = \mathbf{\Phi}$ with $\mathbf{\Phi}$ being the factor correlation.  In this case,  the covariance matrix of the observed variables $\bm{X}$ is expressed as $ \mathbf{\Lambda} \mathbf{\Phi}\mathbf{\Lambda}^T+\mathbf{\Psi} $.  If each factor is highly correlated to each other, i.e., the absolute values non-diagonal elements of $\mathbf{\Phi}$ are large,  the lasso based on the orthogonal model can often estimate a completely different model from that with the true factor structure.    
A typical example of this phenomena is given as follows:
\begin{Example}\label{orthogonaldoesnotwork}
Suppose that the true factor loadings, unique variances  and factor correlation are given by
\begin{equation*}
\mathbf{\Lambda}=
\left(
\begin{array}{rrrrrr}
0.9&0.9&0.9&0.0&0.0&0.0\\
0.0&0.0&0.0&0.9&0.9&0.9\\
\end{array}
\right)^T, \ \mathbf{\Psi} =0.19\mathbf{I}_6, \
 \mathbf{\Phi} = 
\left(
\begin{array}{rrrrrr}
1.0&0.6\\
0.6&1.0\\
\end{array}
\right).
\label{pss00}
\end{equation*}
We generated 50 observations from $\bm{X} \sim N_6(\mathbf{0}, \mathbf{\Lambda} \mathbf{\Phi} \mathbf{\Lambda}^T + \mathbf{\Psi})$.   The model was estimated by the penalized likelihood method with $P(\theta)=\theta$ (i.e., the lasso) and $\rho=0.01$.  The estimated factor landings were 
\begin{equation}
\hat{\mathbf{\Lambda}}_{\rm ort}=
\left(
\begin{array}{rrrrrr}
0.87 & 0.83 & 0.89 & 0.59 & 0.53 & 0.54 \\ 
0.00 & 0.05 & -0.02 & 0.71 & 0.62 & 0.62 \\ 
\end{array}
\right)^T. \label{orthogonal_eq}
\end{equation}
Because the lasso tends to produce some of the loadings being zero,  $\hat{\lambda}_{12}$, $\hat{\lambda}_{22}$ and $\hat{\lambda}_{32}$ were close to or exactly zero.  However,  $\hat{\lambda}_{41}$, $\hat{\lambda}_{51}$ and $\hat{\lambda}_{61}$  were far from zero although the true parameters are zero.    The lasso based on the orthogonal model was not able to approximate the true factor structure.

\end{Example}

The problem that the lasso based on orthogonal model cannot approximate the true factor structure is closely related to the rotation problem. In orthogonal model,   the true covariance matrix  $\mathbf{\Lambda}\mathbf{\Phi}\mathbf{\Lambda}^T + \mathbf{\Psi}$ is estimated by $\hat{\mathbf{\Lambda}}_{\rm ort}\hat{\mathbf{\Lambda}}_{\rm ort}^T + \hat{\mathbf{\Psi}}_{\rm ort}$, which means $\hat{\mathbf{\Lambda}}_{\rm ort}$ approximates $\mathbf{\Lambda}\mathbf{G}$, where $\mathbf{G}=(g_{ii'})$ is an $m \times m$ matrix that satisfies $\mathbf{G}\mathbf{G}^T=\mathbf{\Phi}$.  The matrix $\mathbf{G}$ is not an identity matrix unless the factor correlation $\mathbf{\Phi}$ is an identity matrix, so that $\hat{\mathbf{\Lambda}}_{\rm ort}$ is not always close to $\mathbf{\Lambda}$ even if $\hat{\mathbf{\Lambda}}_{\rm ort} \approx \mathbf{\Lambda}\mathbf{G}$.  


Furthermore, the matrix $\mathbf{G}$ can have a rotational indeterminacy, since $\mathbf{G}\mathbf{G}^T = \mathbf{G}^*(\mathbf{G}^{*})^T = \mathbf{\Phi}$, where $\mathbf{G}^*=\mathbf{G}\mathbf{T}$ with $\mathbf{T}$ being an arbitrary orthogonal matrix. 
The rotational indeterminacy of $\mathbf{G}$ leads to $\ell(\mathbf{\Lambda}{\mathbf{G}},\mathbf{\Psi})  = \ell(\mathbf{\Lambda}{\mathbf{G}^*},\mathbf{\Psi}) $.  Thus, if $\hat{\mathbf{\Lambda}}_{\rm ort}$ and $\hat{\mathbf{\Psi}}_{\rm ort}$ are expressed as $\hat{\mathbf{\Lambda}}_{\rm ort} = \mathbf{\Lambda}{\mathbf{G}}$ and $\hat{\mathbf{\Psi}}_{\rm ort}=\mathbf{\Psi}$, the matrix $\mathbf{G} $ is obtained by solving the following problem:
\begin{equation*}
\max_{\mathbf{G}} \ell_{\rho}^{\rm ort}(\mathbf{\Lambda}{\mathbf{G}},\mathbf{\Psi})  \quad {\rm s.t.,} \quad \mathbf{G}\mathbf{G}^T=\mathbf{\Phi},
\end{equation*} 
which is equivalent to 
\begin{equation}
\min_{\mathbf{G}} \sum_{i=1}^p\sum_{j=1}^mP(|\breve{\lambda}_{ij}|) \quad {\rm s.t.,} \quad \mathbf{G}\mathbf{G}^T=\mathbf{\Phi},\label{rotationproblem0}
\end{equation} 
where $\breve{\lambda}_{ij}$ is the $(i,j)$-th element of $\mathbf{\Lambda}{\mathbf{G}}$.

How is the matrix $\mathbf{G}$ estimated?  To explain this, we assume that the true factor loadings $\mathbf{\Lambda}$ possess perfect simple structure, that is, each row has at most one nonzero element.  
The problem in (\ref{rotationproblem0}) is then written as 
\begin{equation}
\min_{\mathbf{G}}\sum_{i=1}^m \sum_{i'=1}^m P(w_{ii'}|g_{ii'}|) \quad {\rm s.t.,} \quad \mathbf{G}\mathbf{G}^T=\mathbf{\Phi},\label{rotationproblem2}
\end{equation} 
where $w_{ii'}$ are positive values. Because the objective function is based on the $L_1$ loss, some of the elements of $\mathbf{G}$ become exactly zero.  
Empirically, one of the elements of $\mathbf{G}$ often becomes 1. When $g_{qr}=1$, we have $g_{qi'}=0$ $(i'\ne r)$ and $g_{ir} = \phi_{ir}$ $(i \ne q)$, so that  all elements of $r$-th column of $\mathbf{\Lambda}\mathbf{G}$ become non-zero unless $\phi_{ir} =  0$, which does not approximate the perfect simple structure.   In this way, the lasso-type penalization via orthogonal model can often estimate a completely different model from that with the true factor structure when the common factors are  highly correlated.  

\begin{Example}
In Example \ref{orthogonaldoesnotwork}, the problem in (\ref{rotationproblem2}) is written as
\begin{equation*}
\min_{\mathbf{G}} (|g_{11}| + |g_{12}| + |g_{21}| + |g_{22}|) \quad {\rm s.t.,} \quad \mathbf{G}\mathbf{G}^T=\left(
\begin{array}{rrrrrr}
1.0&0.6\\
0.6&1.0\\
\end{array}
\right).
\label{rotationproblem2example2}
\end{equation*} 
The solution of $\mathbf{G}$ is given by 
$$
\mathbf{G} = 
\begin{pmatrix}
1.0 & 0.0 \\ 
0.6 & 0.8 \\ 
\end{pmatrix}
.$$
In this case, we have
\begin{equation*}
\mathbf{\Lambda}{\mathbf{G}} = \left(
\begin{array}{rrrrrr}
0.90 & 0.90 & 0.90 & 0.54 & 0.54 & 0.54 \\ 
0.00 & 0.00 & 0.00 & 0.72 & 0.72 & 0.72 \\ 
\end{array}
\right)^T, \label{orthogonal_eq_true}
\end{equation*} 
which is quite similar to the maximum penalized likelihood estimates of factor loadings based on orthogonal model in (\ref{orthogonal_eq}). 
\end{Example}

\section{Estimation of oblique structure via penalized likelihood factor analysis}
The lasso-type penalized likelihood factor analysis based on the orthogonal model cannot often approximate the oblique structure as described in the Section \ref{sec:Problem of orthogonal model}.  In this Section, we propose to incorporate a factor correlation into the model, and estimate the oblique model by maximum penalized likelihood procedure.  
\subsection{Model Estimation}
Let $\ell(\mathbf{\Lambda},\mathbf{\Psi},\mathbf{\Phi})$ be the log-likelihood function based on the oblique model
\begin{eqnarray}
\ell(\mathbf{\Lambda}} ,{\mathbf{\Psi},\mathbf{\Phi})= -\frac{N}{2} \left[ p\log(2\pi)+\log |\mathbf{\Lambda} \mathbf{\Phi}\mathbf{\Lambda}^T+\mathbf{\Psi}| + \mathrm{tr} \{ (\mathbf{\Lambda} \mathbf{\Phi} \mathbf{\Lambda}^T+\mathbf{\Psi})^{-1} \mathbf{S} \} \right],\label{taisuuyuudo}
\end{eqnarray}
and $\ell_{\rho}(\mathbf{\Lambda},\mathbf{\Psi},\mathbf{\Phi})$ be the penalized log-likelihood function
\begin{equation}
\ell_{\rho}(\mathbf{\Lambda},\mathbf{\Psi},\mathbf{\Phi})  =  \ell(\mathbf{\Lambda},\mathbf{\Psi},\mathbf{\Phi})  -N   \sum_{i=1}^p\sum_{j=1}^m \rho P(|\lambda_{ij}|).  \label{problem_rotation_pmle3_adj_obl}
\end{equation}
We estimate the factor loadings, unique variance,  and factor correlation, say,  $\hat{\mathbf{\Lambda}}_{\rm obl}$,  $\hat{\mathbf{\Psi}}_{\rm obl}$,  and $\hat{\mathbf{\Phi}}_{\rm obl}$ (``obl" is an abbreviation for oblique),   by maximum  penalized likelihood procedure simultaneously:
\begin{equation*}
(\hat{\mathbf{\Lambda}}_{\rm obl},  \hat{\mathbf{\Psi}}_{\rm obl}, \hat{\mathbf{\Phi}}_{\rm obl}) = \mathrm{arg}\max_{\mathbf{\Lambda},\mathbf{\Psi},\mathbf{\Phi}}\ell_{\rho}(\mathbf{\Lambda},\mathbf{\Psi},\mathbf{\Phi}).\label{problem_rotation_pmle3_adj0_obl}
\end{equation*}

\begin{Example}
The lasso based on oblique model was applied to the dataset used in  the Example \ref{orthogonaldoesnotwork}.  
When $\rho=0.01$, the estimates of factor loadings 
were
\begin{equation*}
\hat{\mathbf{\Lambda}}_{\rm obl}=
\left(
\begin{array}{rrrrrr}
0.85 & 0.78 & 0.89 & 0.00 & 0.00 & 0.02 \\ 
0.03 & 0.09 & 0.00 & 0.93 & 0.82 & 0.81 \\ 
\end{array}
\right)^T, 
\end{equation*}
which closely approximate the true factor loadings compared with orthogonal factor loadings $\hat{\mathbf{\Lambda}}_{\rm ort }$ in (\ref{orthogonal_eq}).  
\end{Example}

\subsection{Algorithm}
It is well known that the solutions estimated by the lasso-type regularization methods are not usually expressed in a closed form mainly because the penalty term includes a nondifferentiable function.    In regression analysis, a number of researchers have proposed fast algorithms to obtain the entire solutions (e.g., Least angle regression, \citealp{Efronetal:2004}; Coordinate descent algorithm, \citealp{Friedmanetal:2007}; Generalized path seeking, \citealp{Friedman2008}).   
In particular, the coordinate  descent algorithm is known as a remarkably fast algorithm \citep{Friedmanetal:2010} and can also  be applied to a wide variety of convex and nonconvex penalties \citep{breheny2011coordinate,Mazumderetal:2009}.  Thus, we employ the coordinate descent algorithm to obtain the entire solution.  

In the coordinate descent algorithm, each step is fast if an explicit formula for each coordinate-wise maximization is given,  whereas the log-likelihood function in (\ref{taisuuyuudo}) may not lead to the explicit formula.    
In order to derive the explicit formula, we apply the EM algorithm  \citep{rubin1982algorithms} to the penalized likelihood factor analysis.  The coordinate descent algorithm is utilized to maximize the nonconcave function in the maximization step of the EM algorithm.  Because the complete-data log-likelihood function takes the quadratic form, the explicit formula for each coordinate-wise maximization is available. 

\subsubsection{Update Equation for Fixed Regularization Parameter}
First, we provide the update equations of factor loadings, unique variances, and factor correlation when $\rho$ and $\gamma$ are fixed.  Suppose that $\mathbf{\Lambda}_{\rm old}$, $\mathbf{\Psi}_{\rm old}$ and $\mathbf{\Phi}_{\rm old}$  are the current values of factor loadings, unique variances, and factor correlation.   The model can be estimated by maximizing the expectation of the complete-data penalized log-likelihood function with respect to $\mathbf{\Lambda}$, $\mathbf{\Psi}$ and $\mathbf{\Phi}$:
\begin{eqnarray}
E[\l_{\rho}^{C} (\mathbf{\Lambda},\mathbf{\Psi},\mathbf{\Phi})] &=&- \frac{N}{2} \sum_{i=1}^p \log \psi_i - \frac{N}{2} \sum_{i=1}^p \frac{s_{ii} - 2\bm{\lambda}_i^T\mathbf{b}_i+ \bm{\lambda}_i^T \mathbf{A}\bm{\lambda}_i}{\psi_i}
 \cr
&&- \frac{N}{2} \log |\mathbf{\Phi}| - \frac{N}{2} {\rm  tr} ( \mathbf{\Phi}^{-1} \mathbf{A}) - N   \sum_{i=1}^p\sum_{j=1}^m \rho P(|\lambda_{ij}|) + {\rm const.},\cr
&& \label{ECL}
\end{eqnarray}
where $\mathbf{b}_i = \mathbf{M}^{-1}\mathbf{\Lambda}_{\rm old}^T\mathbf{\Psi}_{\rm old }^{-1}\mathbf{s}_i$ and $\mathbf{A}=\mathbf{M} ^{-1} + \mathbf{M}^{-1}\mathbf{\Lambda}_{\rm old}^T\mathbf{\Psi}_{\rm old }^{-1}\mathbf{S}\mathbf{\Psi}_{\rm old }^{-1}\mathbf{\Lambda}_{\rm old }\mathbf{M}^{-1}$. Here $\mathbf{M} = \mathbf{\Lambda}_{\rm old}^T\mathbf{\Psi}_{\rm old}^{-1}\mathbf{\Lambda}_{\rm old} + \mathbf{\Phi}^{-1}_{\rm old}$, and $\mathbf{s}_i$ is the $i$-th column vector of $\mathbf{S}$.  The derivation of the complete-data penalized log-likelihood function is described in Appendix. 

The new parameter $(\mathbf{\Lambda}_{\rm new}, \mathbf{\Psi}_{\rm new},\mathbf{\Phi}_{\rm new})$ can be computed by maximizing the complete-data penalized log-likelihood function, i.e., 
\begin{eqnarray}
(\mathbf{\Lambda}_{\rm new}, \mathbf{\Psi}_{\rm new},\mathbf{\Phi}_{\rm new}) = {\rm arg}\max_{\mathbf{\Lambda}, \mathbf{\Psi},\mathbf{\Phi}} E[\l_{\rho}^{C} (\mathbf{\Lambda},\mathbf{\Psi},\mathbf{\Phi})] .\label{maxECL}
\end{eqnarray}
The solution in (\ref{maxECL}) is not usually expressed in a closed form because the penalty term includes a nondifferentiable function, so that the coordinate descent algorithm is utilized.


Let  $\tilde{\bm{\lambda}}_{i}^{(j)} $ be an ($m-1$)-dimensional vector  $( \tilde{\lambda}_{i1},\tilde{\lambda}_{i2},\dots,\tilde{\lambda}_{i(j-1)},\tilde{\lambda}_{i(j+1)},\dots,\tilde{\lambda}_{im})^T$.  The parameter $\lambda_{ij}$ can be updated by maximizing (\ref{ECL}) with the other parameters $\tilde{\bm{\lambda}}_{i}^{(j)} $, $\mathbf{\Psi}$ and $\mathbf{\Phi}$ being fixed, i.e.,  we solve the following problem:
\begin{eqnarray*}
\tilde{\lambda}_{ij} &=& {\rm arg} \min_{\lambda_{ij}}  \frac{1}{2\psi_i} \left\{a_{jj}\lambda_{ij}^2  - 2\left(b_{ij} - \sum_{k \ne j} a_{kj} \tilde{\lambda}_{ik} \right)\lambda_{ij} \right\}  + \rho   P( |\lambda_{ij}|) \cr
&=& {\rm arg} \min_{\lambda_{ij}}\frac{1}{2}   \left( \lambda_{ij}  - \frac{b_{ij} - \sum_{k \ne j} a_{kj} \tilde{\lambda}_{ik} }{a_{jj}} \right)^2  + \frac{\psi_i\rho}{a_{jj}} P( |\lambda_{ij}|). \label{lambdaupdate} 
\end{eqnarray*}
This is equivalent to minimizing the following penalized squared-error loss function 
\begin{equation*}
S(\tilde{\theta}) = {\rm arg} \min_{\theta} \left\{ \frac{1}{2}(\theta - \tilde{\theta})^2 + \rho^* P(|\theta|) \right\}. \label{lamdba_update_CD}
\end{equation*}
The solution $S(\tilde{\theta}) $ can be expressed in a closed form for a variety of convex and nonconvex penalties.  For example, the update equation for MC+ penalty is given by 
\begin{equation*}
S(\tilde{\theta})=\left\{
\begin{array}{ll}
 \dfrac{{\rm sgn}(\tilde{\theta})(|\tilde{\theta}| - \rho^*)_+}{1-1/\gamma} & \mbox{if $ |\tilde{\theta}| \le \rho^*\gamma$}\\
\tilde{\theta} & \mbox{if $|\tilde{\theta}| > \rho^*\gamma$}.\\
\end{array}
 \right.
 \label{MCsolution}
\end{equation*}


After updating $\mathbf{\Lambda}$ by the coordinate descent algorithm, the new values of $\mathbf{\Psi}_{\rm new}$ and $\mathbf{\Phi}_{\rm new}$ are obtained by maximizing the expected penalized log-likelihood function in (\ref{ECL}) as follows:
\begin{eqnarray*}
(\psi_i )_{\rm new} &=& s_{ii} - 2 (\hat{\bm{\lambda}}^T_i)_{\mathrm{new}}\mathbf{b}_i  +  (\hat{\bm{\lambda}}_i)_{\mathrm{new}}^T \mathbf{A} (\hat{\bm{\lambda}}_i)_{\mathrm{new}} \quad  \mbox{for $i=1,\dots,p$,}\\
\mathbf{\Phi}_{\rm new} &=& {\rm arg} \min_{\mathbf{\Phi}} \{ \log |\mathbf{\Phi}| + {\rm tr}(\mathbf{\Phi}^{-1}\mathbf{A}) \},
\end{eqnarray*}
where $(\psi_i )_{\rm new} $ is the $i$-th diagonal element of $\mathbf{\Psi}_{\rm new}$ and $(\hat{\bm{\lambda}}_i)_{\mathrm{new}}$ is the $i$-th column of $\hat{\mathbf{\Lambda}}_{\rm new}$. The explicit formula of  $\mathbf{\Phi}_{\rm new} $ may not be easily derived, because the diagonal elements of $\mathbf{\Phi}$ are fixed by 1.  Therefore, the non-diagonal elements of $\mathbf{\Phi}_{\rm new}$ are estimated by Broyden-Fletcher-Goldfarb-Shanno (BFGS) optimization procedure.  

\subsubsection{Pathwise Algorithm}
A pathwise algorithm for orthogonal case has been proposed by \citet{hirose2012sparse}, and we apply their algorithm to the oblique case.  The pathwise algorithm can produce the solution for the grid of  increasing $\rho$ values $P=\{\rho_1,\dots,\rho_K  \}$ and a grid of increasing values $\Gamma=\{ \gamma_1,\dots,\gamma_T \}$ efficiently, where $\gamma_T$ gives the lasso penalty (e.g., $\gamma_T=\infty$ for MC+ family).     
First, we compute the lasso solution path for  $P=\{\rho_1,\dots,\rho_K  \}$ by decreasing the sequence of values for $\rho$, starting with the largest value $\rho=\rho_K$ for which the estimates of factor loadings $\hat{\mathbf{\Lambda}}=\mathbf{O}$.  
Next, the value of $\gamma_{T-1}$ is selected, and the solutions are produced for the sequence of $P=\{\rho_1,\dots,\rho_K  \}$.  
The  solution at ($\gamma_{T-1},\rho_k$) can be computed by using the solution at  ($\gamma_{T},\rho_k$), which leads to improved and smoother objective value surfaces \citep{Mazumderetal:2009}. In the same way, for $t=T-2,\dots,1$, the solution at ($\gamma_{t},\rho_k$) can be computed by using the solution at  ($\gamma_{t+1},\rho_k$).  

\subsection{Selection of the Regularization Parameter}
In this modeling procedure, it is important to select the appropriate value of the regularization parameter $\rho$.  The following two selection procedures are introduced.  

\subsubsection{Model Selection Criteria}\label{sec:MSC}
 The selection of the regularization parameter can be viewed as a model selection and evaluation problem. In regression analysis, the degrees of freedom of the lasso \citep{Zouetal:2007} may be used for selecting the regularization parameter.  With the use of the degrees of freedom, the following model selection criteria are introduced:
\begin{eqnarray*}
{\rm AIC} &=& -2 \ell(\hat{\mathbf{\Lambda}} ,\hat{\mathbf{\Psi}}, \hat{\mathbf{\Phi}}) + 2 p^*\\
{\rm BIC} &=& -2 \ell(\hat{\mathbf{\Lambda}} ,\hat{\mathbf{\Psi}},  \hat{\mathbf{\Phi}}) + p^* \log N  ,\\
{\rm CAIC} &=& -2 \ell(\hat{\mathbf{\Lambda}} ,\hat{\mathbf{\Psi}}, \hat{\mathbf{\Phi}}) +  p^*(\log N + 1) ,
\end{eqnarray*}
where  the number of parameters is given by $p^*=df(\rho_k) + m_0(m_0-1) /2 + p$.  Here $df(\rho_k)$ is the number of nonzero parameters for the lasso penalty at $\rho=\rho_k$, $m_0(m_0-1) /2$ is the number of parameters in factor correlation matrix and $p$ is the number of parameters in unique variances.  Note that this formula can be applied to any value of $\gamma$ if the reparameterization of the penalty function \citep{Mazumderetal:2009} is carried out, because the  reparameterization constrains the degrees of freedom to be constant as $\gamma$ varies.  

\subsubsection{Goodness-of-Fit Index}
It may be easy to interpret the estimated model when the factor loadings are sufficiently sparse.  However, a model that is too sparse does not fit the data.  Therefore, it is reasonable to select a regularization parameter that produces sparse solutions and also yields large values for the following goodness-of-fit index (GFI) and the adjusted GFI (AGFI):
\begin{eqnarray*}
{\rm GFI} &=& 1- \frac{ {\rm tr} [ \{ \hat{\mathbf{\Sigma}}^{-1}  (\mathbf{S} - \hat{\mathbf{\Sigma}} ) \}^2 ] }{  {\rm tr} [ \{ \hat{\mathbf{\Sigma}}^{-1} \mathbf{S} \}^2 ] }, \\
{\rm AGFI} &=& 1-\frac{p(p+1)(1-{\rm GFI})}{p(p+1)-2df},
\end{eqnarray*}
 where  $\hat{\mathbf{\Sigma}} = \hat{\mathbf{\Lambda}} \hat{\mathbf{\Phi}}\hat{\mathbf{\Lambda}}^T  + \hat{\mathbf{\Psi}} $.  The GFI and AGFI take values from 0 through 1.  In our experience, the model is fitted well if the value of the GFI is greater than 0.9.   
 
 \subsection{Treatment for Improper Solutions}
 It is well-known that the maximum likelihood estimates of unique variances can turn out to be zero or negative, which is referred as the improper solutions, and many researchers have studied this problem (e.g., \citealp{van1978various,anderson1984effect,kano1998improper}).  In general, the occurrence of improper solutions makes converge of the algorithm slow and unstable.  In order to handle this issue, we add a penalty with respect to $\mathbf{\Psi}$ to (\ref{problem_rotation_pmle3_adj_obl}) according to the basic idea given by \citet{martin1975bayesian} and \citet{hirose2011bayesian}: 
\begin{equation*}
\ell_{\rho}^*(\mathbf{\Lambda},\mathbf{\Psi},\mathbf{\Phi})  =  \ell_{\rho}(\mathbf{\Lambda},\mathbf{\Psi},\mathbf{\Phi})   -\frac{N}{2} \eta{\rm tr}(\mathbf{\Psi}^{-1/2}\mathbf{S}\mathbf{\Psi}^{-1/2}),  \label{problem_rotation_pmle3_adj2}
\end{equation*}
where $\eta$ is a tuning parameter.  Note that when  $\psi_i \rightarrow 0$, ${\rm tr}(\mathbf{\Psi}^{-1/2}\mathbf{S}\mathbf{\Psi}^{-1/2}) \rightarrow \infty$.  Thus, the penalty term ${\rm tr}(\mathbf{\Psi}^{-1/2}\mathbf{S}\mathbf{\Psi}^{-1/2})$ prevents the occurrence of improper solutions.  \citet{hirose2011bayesian} derived a generalized Bayesian information criterion \citep{konishi2004bayesian} for selecting the appropriate value of $\eta$, whereas it is difficult to derive generalized Bayesian model criterion in lasso-type penalization procedure.  In practice, the penalty term can prevent the occurrence of improper solution even when $\eta$ is very small such as 0.001.  

We provide a package {\tt fanc} in {\tt R} \citep{R:2010}, which implements our algorithm to produce the entire solution path.  The package  {\tt fanc} is available from Comprehensive R Archive Network (CRAN) at \url{http://cran.r-project.org/web/packages/fanc/index.html}.  
\section{Numerical Examples}
\subsection{Monte Carlo Simulations}\label{sec:simulation}
In the simulation study, we used three models according to the following factor loadings:
\begin{flushleft}
{\bf Model (A):} 
\end{flushleft}
\vspace{-2em}
\begin{eqnarray*}
\mathbf{\Lambda} &=& 
\left(
\begin{array}{rrrrrr}
0.9&0.9&0.9&0.0&0.0&0.0\\
0.0&0.0&0.0&0.8&0.8&0.8\\
\end{array}
\right)^T, 
\end{eqnarray*}
\begin{flushleft}
{\bf Model (B):} 
\end{flushleft}
\vspace{-2em}
\begin{eqnarray*}
\mathbf{\Lambda} &=& 
\left(
\begin{array}{rrrrrrrrrr}
0.9&0.9&0.9&0.0&0.0&0.0&0.0&0.0&0.0\\
0.0&0.0&0.0&0.8&0.8&0.8&0.0&0.0&0.0\\
0.0&0.0&0.0&0.0&0.0&0.0&0.7&0.7&0.7\\
\end{array}
\right)^T,
\end{eqnarray*}
\begin{flushleft}
{\bf Model (C):} 
\end{flushleft}
\vspace{-1em}
 \begin{eqnarray*}
 \mathbf{\Lambda} &=& 
 \left(
 \begin{array}{cccc}
 0.9 \cdot \mathbf{1}_{25} &\mathbf{0}_{25} & \mathbf{0}_{25} &\mathbf{0}_{25} \\
\mathbf{0}_{25} &0.8\cdot\mathbf{1}_{25} & \mathbf{0}_{25}  & \mathbf{0}_{25}  \\
\mathbf{0}_{25} &\mathbf{0}_{25} &0.7 \cdot\mathbf{1}_{25}  & \mathbf{0}_{25}  \\
\mathbf{0}_{25} &\mathbf{0}_{25} & \mathbf{0}_{25}  & 0.6 \cdot\mathbf{1}_{25}  \\
 \end{array}
 \right),
 \end{eqnarray*}
 where $\mathbf{1}_{q}$ is a $q$-dimensional vector with each element being 1, and $\mathbf{0}_{q}$ is a $q$-dimensional zero vector.  For all models, we set $\mathbf{\Phi} = 0.4 \cdot \mathbf{I}_m + 0.6 \cdot\mathbf{1}_m\mathbf{1}_m^T$, and $\mathbf{\Psi} = {\rm diag}( \mathbf{I}_m - \mathbf{\Lambda} \mathbf{\Phi} \mathbf{\Lambda}^T)$.  The Model (C) is a relatively large model compared with Models (A) and (B).

For each model, 1000 data sets were generated with $\bm{x} \sim N_p(\mathbf{0},\mathbf{\Lambda}\mathbf{\Phi}\mathbf{\Lambda}^T + \mathbf{\Psi})$.     The number of observations was $N=50, 100$, and $200$.  The model was estimated by the maximum penalized likelihood method based on both orthogonal model and oblique model.  The penalty functions were the MC+ family with $\gamma=2.10$ and the lasso, and the regularization parameter was selected by the AIC, BIC and CAIC.  For comparison, we also estimated the model by the maximum likelihood method, and employed the rotation techniques based on the following criteria: the lasso loss criterion (both orthogonal and oblique models), the varimax criterion (orthogonal model) and promax criterion (oblique model).  For example, the lasso loss criterion for oblique model is formulated as follows:
\begin{equation*}
\min_{\mathbf{T}} \sum_{i=1}^p\sum_{j=1}^m|\hat{\lambda}_{ij}^*|, \quad {\rm s.t.} \quad \hat{\mathbf{\Lambda}}^*=\hat{\mathbf{\Lambda}}_{\rm MLE}\mathbf{T}, \ {\rm diag}(\mathbf{T}^T\mathbf{T})=\mathbf{I},
\end{equation*}
where $\hat{\mathbf{\Lambda}}^*=(\hat{\lambda}_{ij}^*)$ and $\hat{\mathbf{\Lambda}}_{\rm MLE}$ is the maximum likelihood estimates of factor loadings.  Note that  the lasso loss function is included in the class of component loss function \citep{jennrich2004rotation,jennrich2006rotation}.  

Tables \ref{table:simulation1}, \ref{table:simulation2} and \ref{table:simulation3} show the mean squared error of factor loadings,  
the true positive rate (TPR), and true negative rate (TNR).   The mean squared error is defined by ${\rm MSE}=\sum_{s=1}^{1000} \| \mathbf{\Lambda} - \hat{\mathbf{\Lambda}}^{(s)} \|^2 / (1000pm) $, where $\hat{ \mathbf{\Lambda}}^{(s)} $ is the estimated factor loading for the $s$-th dataset.   The TPR (TNR)  indicates the proportion of cases where non-zero (zero) factor loadings correctly set to non-zero (zero).  Note that the maximum likelihood estimates are not available when $N \le p$, so that the results of rotation techniques based on the maximum likelihood estimates for $N=50$ and $N=100$ in Model (C) were not displayed. 
We can see that 
\begin{itemize}
\item The lasso-type regularization with orthogonal model yielded large MSE and small TNR  even when  the number of observations $N$ was sufficiently large, which suggests  the orthogonal model may produce different factor structure from the true one.    
\item For Model (C), although the maximum likelihood estimates were not available when $N=50$ and $N=100$, the penalized likelihood procedure via BIC and CAIC with MC+ relatively selected correct models.  
\item The MC+ family often performed better than the lasso in terms of both the MSE and model consistency.  
\item The BIC and CAIC may often select the correct model compared with the AIC.  
\end{itemize}

\begin{table}[!t]
\caption{Mean squared error, the true positive rate (TPR), and true negative rate (TNR) for Model (A). In the second column, ``obl" and ``ort" indicate the oblique model and orthogonal model, respectively.   In the last column, ``P/V"  indicates the promax rotation for oblique case, and the varimax rotation for orthogonal case.} \label{table:simulation1}
\begin{center}
\begin{tabular}{llrrrrrrrrrr}
  \hline
&& & \multicolumn{6}{c}{Penalization}&  \multicolumn{2}{c}{Rotation } \\ 
& && \multicolumn{2}{c}{AIC}&  \multicolumn{2}{c}{BIC}&  \multicolumn{2}{c}{CAIC} &---&---  \\ 
$N$&  && MC+ & lasso   & MC+ & lasso   & MC+ & lasso & lasso & P/V\\ 
  \hline
$50$ &obl &MSE &         0.14 & 0.18 & 0.14 & 0.20 & 0.16 & 0.24 & 0.17 & 0.14 \\
&&                            TPR&   1.00 & 1.00 & 1.00 & 1.00 & 0.99 & 1.00 & 1.00 & 1.00 \\
&&                            TNR &  0.75 & 0.47 & 0.84 & 0.55 & 0.86 & 0.57 & 0.05 & 0.00 \\
&ort &MSE&             1.21 & 1.13 & 1.21 & 1.08 & 1.21 & 1.02 & 1.27 & 0.53\\ 
&&  TPR&                             0.97 & 0.98 & 0.97 & 0.98 & 0.96 & 0.98 & 0.98 & 1.00 \\
&&  TNR &                            0.30 & 0.20 & 0.33 & 0.23 & 0.35 & 0.26 & 0.14 & 0.00 \\
   \hline
$100$ &obl &MSE &  0.06 & 0.09 & 0.04 & 0.10 & 0.03 & 0.11 & 0.08 & 0.06 \\ 
&&  TPR&                        1.00 & 1.00 & 1.00 & 1.00 & 1.00 & 1.00 & 1.00 & 1.00 \\ 
&&  TNR  &                      0.81 & 0.46 & 0.91 & 0.56 & 0.93 & 0.58 & 0.06 & 0.00 \\ 
&ort  &MSE&        1.03 & 0.95 & 1.05 & 0.90 & 1.05 & 0.87 & 1.04 & 0.48 \\ 
&&  TPR&                        0.99 & 0.99 & 0.98 & 0.99 & 0.98 & 0.99 & 0.99 & 1.00 \\ 
&&  TNR &                       0.34 & 0.21 & 0.35 & 0.24 & 0.36 & 0.25 & 0.14 & 0.00 \\ 
   \hline
$200$ &obl &MSE &   0.02 & 0.05 & 0.01 & 0.06 & 0.01 & 0.06 & 0.04 & 0.03 \\ 
&&  TPR&                         1.00 & 1.00 & 1.00 & 1.00 & 1.00 & 1.00 & 1.00 & 1.00 \\ 
&&  TNR  &                       0.87 & 0.47 & 0.97 & 0.56 & 0.97 & 0.60 & 0.07 & 0.00 \\ 
&ort &MSE&         0.89 & 0.84 & 0.89 & 0.78 & 0.89 & 0.77 & 0.88 & 0.46 \\ 
&&  TPR&                         0.99 & 1.00 & 0.99 & 1.00 & 0.99 & 1.00 & 1.00 & 1.00 \\ 
&&  TNR &                        0.41 & 0.21 & 0.42 & 0.26 & 0.43 & 0.26 & 0.14 & 0.00 \\ 
\hline
\end{tabular}
\end{center}
\end{table}

\begin{table}[!t]
\caption{Mean squared error, the true positive rate (TPR), and true negative rate (TNR) for Model  (B).  In the second column, ``obl" and ``ort" indicate the oblique model and orthogonal model, respectively. In the last column, ``P/V"  indicates the promax rotation for oblique case, and the varimax rotation for orthogonal case.} \label{table:simulation2}
\begin{center}
\begin{tabular}{llrrrrrrrrrr}
  \hline
& && \multicolumn{6}{c}{Penalization}&  \multicolumn{2}{c}{Rotation } \\ 
 &&& \multicolumn{2}{c}{AIC}&  \multicolumn{2}{c}{BIC}&  \multicolumn{2}{c}{CAIC} &---&---  \\ 
$N$&  && MC+ & lasso   & MC+ & lasso   & MC+ & lasso & lasso & P/V\\ 
  \hline
$50$ &obl &MSE &         0.84 & 0.81 & 1.00 & 1.21 & 1.08 & 1.41 & 0.81 & 0.72 \\
&&                            TPR&    0.97 & 0.96 & 0.91 & 0.86 & 0.88 & 0.82 & 1.00 & 1.00\\
&&                            TNR &   0.69 & 0.47 & 0.81 & 0.56 & 0.85 & 0.59 & 0.02 & 0.00\\
&ort &MSE&              2.56 & 2.03 & 2.66 & 1.95 & 2.66 & 2.05 & 2.33 & 1.33\\
&&  TPR&                              0.94 & 0.97 & 0.91 & 0.90 & 0.89 & 0.84 & 0.99 & 1.00\\
&&  TNR &                             0.35 & 0.25 & 0.40 & 0.35 & 0.45 & 0.40 & 0.14 & 0.00\\
   \hline
$100$ &obl &MSE &  0.29 & 0.34 & 0.44 & 0.62 & 0.53 & 0.83 & 0.41 & 0.31 \\ 
&&  TPR&                        1.00 & 1.00 & 0.96 & 0.94 & 0.94 & 0.90 & 1.00 & 1.00 \\ 
&&  TNR  &                      0.77 & 0.47 & 0.88 & 0.57 & 0.90 & 0.59 & 0.03 & 0.00 \\ 
&ort &MSE&        2.35 & 2.02 & 2.43 & 1.80 & 2.43 & 1.76 & 2.24 & 1.13 \\ 
&&  TPR&                        0.96 & 0.98 & 0.94 & 0.96 & 0.94 & 0.94 & 0.99 & 1.00 \\ 
&&  TNR &                       0.38 & 0.23 & 0.42 & 0.30 & 0.44 & 0.34 & 0.16 & 0.00 \\ 
   \hline
$200$ &obl &MSE &   0.10 & 0.16 & 0.07 & 0.18 & 0.08 & 0.22 & 0.19 & 0.12 \\ 
&&  TPR&                         1.00 & 1.00 & 1.00 & 1.00 & 0.99 & 0.99 & 1.00 & 1.00 \\ 
&&  TNR  &                       0.86 & 0.48 & 0.95 & 0.56 & 0.96 & 0.58 & 0.03 & 0.00 \\ 
&ort &MSE&         2.05 & 1.85 & 2.11 & 1.64 & 2.13 & 1.59 & 1.95 & 1.02 \\ 
&&  TPR&                         0.98 & 0.99 & 0.97 & 0.99 & 0.97 & 0.99 & 0.99 & 1.00 \\ 
&&  TNR &                        0.41 & 0.21 & 0.44 & 0.27 & 0.45 & 0.28 & 0.16 & 0.00 \\ 
\hline
\end{tabular}
\end{center}
\end{table}

\begin{table}[!t]
\caption{Mean squared error, the true positive rate (TPR), and true negative rate (TNR) for Model  (C). In the second column, ``obl" and ``ort" indicate the oblique model and orthogonal model, respectively. In the last column, ``P/V"  indicates the promax rotation for oblique case, and the varimax rotation for orthogonal case.} \label{table:simulation3}
\begin{center}
\begin{tabular}{llrrrrrrrrrr}
  \hline
& && \multicolumn{6}{c}{Penalization}&  \multicolumn{2}{c}{Rotation} \\ 
 &&& \multicolumn{2}{c}{AIC}&  \multicolumn{2}{c}{BIC}&  \multicolumn{2}{c}{CAIC} &---&---  \\ 
$N$&  && MC+ & lasso   & MC+ & lasso   & MC+ & lasso & lasso & P/V\\ 
  \hline
$50$ &obl &MSE &   8.54 & 10.0 & 7.68 & 17.2 & 8.84 & 20.6 & --- & --- \\ 
&&                            TPR& 0.99 & 1.00 & 0.92 & 0.97 & 0.86 & 0.92 & --- & --- \\ 
&&                            TNR &0.43 & 0.14 & 0.85 & 0.36 & 0.96 & 0.44 & --- & --- \\ 
&ort &MSE&        28.0 & 15.3 & 27.4 & 15.5 & 22.9 & 20.3 & --- & --- \\ 
&&  TPR&                           0.98 & 1.00 & 0.97 & 0.99 & 0.94 & 0.92 & --- & --- \\ 
&&  TNR &                         0.26 & 0.06 & 0.32 & 0.20 & 0.51 & 0.36 & --- & --- \\  
   \hline
$100$ &obl &MSE &3.51 & 5.73 & 1.79 & 12.4 & 2.08 & 13.2 & --- & --- \\ 
&&  TPR&                          1.00 & 1.00 & 0.99 & 1.00 & 0.98 & 0.99 & --- & --- \\ 
&&  TNR  &                        0.58 & 0.14 & 0.99 & 0.41 & 0.99 & 0.43 & --- & --- \\ 
&ort &MSE&        29.5 & 15.5 & 23.0 & 11.9 & 13.8 & 12.6 & --- & --- \\ 
&&  TPR&                          0.98 & 1.00 & 0.98 & 1.00 & 0.97 & 0.99 & --- & --- \\ 
&&  TNR &                         0.32 & 0.04 & 0.51 & 0.13 & 0.74 & 0.22 & --- & --- \\ 
   \hline
$200$ &obl &MSE & 0.97 & 3.39 & 0.67 & 9.24 & 0.67 & 10.2 & 2.18 & 1.68 \\ 
&&  TPR&                           1.00 & 1.00 & 1.00 & 1.00 & 1.00 & 1.00 & 1.00 & 1.00 \\ 
&&  TNR  &                         0.91 & 0.13 & 1.00 & 0.44 & 1.00 & 0.48 & 0.00 & 0.00 \\ 
&ort &MSE&         30.2 & 16.1& 19.2 & 12.1 & 9.12 & 11.1 & 18.0 & 13.0 \\ 
&&  TPR&                           0.99 & 1.00 & 0.99 & 1.00 & 0.99 & 1.00 & 1.00 & 1.00 \\ 
&&  TNR &                          0.41 & 0.03 & 0.65 & 0.07 & 0.86 & 0.09 & 0.02 & 0.00 \\ 
\hline
\end{tabular}
\end{center}
\end{table}

\subsection{Analysis of Harman's psychological tests data} \label{sec:Harman}
We illustrate the proposed procedure by Harman's psychological tests data  \citep{harman1976modern}. This data represents scores of $N =145$ subjects on the  24 psychological tests.  The dataset is available from the {\tt datasets}  in the software {\tt R} \citep{R:2010}.  Table \ref{tab:Harman} shows the  factor loadings estimated by MC+ based on both orthogonal and oblique models at $\gamma=2.10$.   The value of $\rho$ was selected by the BIC.  With the MC+ based on the orthogonal model, all elements of the first column of the estimated factor loadings were relatively large.  This phenomena has been described in Section \ref{sec:Problem of orthogonal model}.  On the other hand, the MC+ based on the oblique model estimated a loading matrix where the first column was sparse.   
The AGFI and GFI of the oblique model were 0.78 and 0.87, respectively, which might be large enough to conclude that  the estimated model fit the observed data.

\begin{table}[!ht]
\caption{Loading matrices estimated by MC+ based on both orthogonal and oblique models at $\gamma=2.10$ for 24 psychological tests data.  The value of $\rho$ was selected by the BIC.}
\label{tab:Harman}
\begin{center}
\begin{tabular}{rrrr|rrrr}
  \hline
\multicolumn{4}{c}{MC+ (Orthogonal)} & \multicolumn{4}{c}{MC+ (Oblique)}\\ 
  \hline
0.75 & $-$0.08 & 0.00 & 0.00 & 0.73 & 0.00 & 0.00 & 0.00 \\ 
0.46 & 0.00 & 0.00 & 0.00 & 0.47 & 0.00 & 0.00 & 0.00 \\ 
 0.56 & 0.00 & 0.16 & 0.00 & 0.66 & 0.00 & $-$0.18 & 0.00 \\ 
0.59 & 0.00 & 0.00 & 0.00 & 0.58 & 0.00 & 0.00 & 0.00 \\ 
 0.47 & 0.63 & $-$0.17 & 0.00 & 0.00 & 0.72 & 0.22 & 0.00 \\ 
0.49 & 0.67 & 0.00 & 0.00 & 0.00 & 0.75 & 0.00 & 0.17 \\ 
0.48 & 0.67 & $-$0.12 & $-$0.12 & 0.00 & 0.79 & 0.13 & 0.00 \\ 
0.56 & 0.41 & $-$0.16 & 0.00 & 0.23 & 0.48 & 0.19 & 0.00 \\ 
 0.49 & 0.71 & 0.00 & 0.00 & 0.00 & 0.81 & 0.00 & 0.13 \\ 
 0.17 & 0.14 & $-$0.81 & 0.24 & $-$0.31 & 0.00 & 0.93 & 0.11 \\ 
0.36 & 0.14 & $-$0.42 & 0.34 & 0.00 & 0.00 & 0.44 & 0.35 \\ 
 0.37 & $-$0.10 & $-$0.63 & 0.13 & 0.14 & $-$0.19 & 0.71 & 0.00 \\ 
0.59 & 0.00 & $-$0.41 & 0.00 & 0.37 & 0.00 & 0.44 & 0.00 \\ 
 0.23 & 0.25 & 0.00 & 0.48 & 0.00 & 0.00 & 0.00 & 0.58 \\ 
 0.26 & 0.15 & 0.00 & 0.45 & 0.00 & 0.00 & 0.00 & 0.53 \\ 
 0.51 & 0.00 & 0.10 & 0.42 & 0.44 & $-$0.14 & $-$0.15 & 0.49 \\ 
0.26 & 0.19 & $-$0.11 & 0.54 & 0.00 & 0.00 & 0.00 & 0.65 \\ 
0.43 & 0.00 & $-$0.19 & 0.43 & 0.28 & $-$0.20 & 0.18 & 0.43 \\ 
0.38 & 0.06 & 0.00 & 0.30 & 0.22 & 0.00 & 0.00 & 0.35 \\ 
 0.56 & 0.25 & 0.00 & 0.16 & 0.37 & 0.24 & 0.00 & 0.17 \\ 
0.55 & 0.00 & $-$0.31 & 0.16 & 0.38 & 0.00 & 0.39 & 0.00 \\ 
0.56 & 0.24 & 0.00 & 0.16 & 0.37 & 0.23 & 0.00 & 0.18 \\ 
0.67 & 0.19 & $-$0.10 & 0.10 & 0.50 & 0.22 & 0.15 & 0.00 \\ 
0.43 & 0.29 & $-$0.41 & 0.24 & 0.00 & 0.22 & 0.47 & 0.22 \\ 
     \hline
\end{tabular}
\end{center}
\end{table}

\section{Concluding  remarks}
In exploratory factor analysis, the lasso based on the orthogonal model often fails in approximating the oblique structure.  We have shown that this disadvantage comes from the rotation problem of factor loadings.  Then, a maximum penalized likelihood factor analysis based on the oblique model has been proposed to handle this problem.  
Our modeling strategy has been investigated through Monte Carlo simulations and the analysis of a real data.    Simulation results show that the proposed procedure can yield much smaller mean squared error and true negative rate  compared with the penalized likelihood factor analysis via the orthogonal model.  Furthermore, the MC+ often produced sparser solutions than the lasso, so that the true factor structure can often be reconstructed.  In the Harman's psychological data example, the orthogonal model estimated factor loadings where first column was dense, whereas our procedure produced sparse factor loadings.

As a future research topic, it would be interesting to construct a penalization procedure via nonconvex penalties for structural equation modeling, such as LISREL \citep{joreskog1996lisrel}, which is able to express much more complex covariance structure between observable variables and common factors. In this paper, the tuning parameter was selected by the information criteria based on the degrees of freedom of the lasso.  The degrees of freedom of the lasso are usually applied to the regression model, whereas we have not given a mathematical support for the degrees of freedom of the lasso in  factor analysis model yet.   Another interesting topic is to provide a theoretical justification of the information criteria given by Section \ref{sec:MSC}.  

\renewcommand{\theequation}{A\arabic{equation}}
\setcounter{equation}{0}
\section*{Appendix: Derivation of complete-data penalized log-likelihood function in EM algorithm}
 In order to apply the EM algorithm, first, the common factors $\bm{f}_n$ can be regarded as missing data and maximize the complete-data penalized log-likelihood function
\begin{equation*}
\l_{\rho}^{C} (\mathbf{\Lambda},\mathbf{\Psi}, \mathbf{\Phi}) = \sum_{n=1}^N \log f(\bm{x}_n,\bm{f}_n) - N   \sum_{i=1}^p\sum_{j=1}^m \rho P(|\lambda_{ij}|), \label{g_C}
\end{equation*}
where the density function $f(\bm{x}_n,\bm{f}_n)$ is defined by
\begin{eqnarray*}
f(\bm{x}_n,\bm{f}_n) &=&(2\pi)^{-p/2} |\mathbf{\Psi}|^{-1/2}  \exp \left\{ - \frac{ (\bm{x}_n - \mathbf{\Lambda}\bm{f}_n)^T  \mathbf{\Psi}^{-1}  (\bm{x}_n - \mathbf{\Lambda}\bm{f}_n) }{2} \right\}       \\
&& \cdot \ (2\pi)^{-m/2}|\mathbf{\Phi}|^{-1/2}  \exp \left( - \frac{ \bm{f}_n^T  \mathbf{\Phi}^{-1}  \bm{f}_n }{2} \right)\\
&=&\prod_{i=1}^p \left\{  (2\pi\psi_i)^{-1/2} \exp \left( - \frac{ (x_{ni}-\bm{\lambda}_i^T\bm{f}_n )^2}{2\psi_i}  \right) \right\} \\
&& \cdot \ (2\pi)^{-m/2}|\mathbf{\Phi}|^{-1/2}  \exp \left( - \frac{ \bm{f}_n^T  \mathbf{\Phi}^{-1}  \bm{f}_n }{2} \right)
\end{eqnarray*}
Then, the expectation of  $\l_{\rho}^{C} $ can be taken with respect to the distributions $f(\bm{f}_n | \bm{x}_n,\mathbf{\Lambda},\mathbf{\Psi},\mathbf{\Phi})$,
\begin{eqnarray*}
E[\l_{\rho}^{C} (\mathbf{\Lambda},\mathbf{\Psi}, \mathbf{\Phi})] &=&-\frac{N(p+m)}{2} \log(2\pi) - \frac{N}{2} \sum_{i=1}^p \log \psi_i \\
&&- \frac{1}{2} \sum_{n=1}^N\sum_{i=1}^p \frac{x_{ni}^2 - 2x_{ni}\bm{\lambda}_i^TE[\bm{F}_n|\bm{x}_n]+ \bm{\lambda}_i^T E[\bm{F}_n\bm{F}_n^T|\bm{x}_n]\bm{\lambda}_i}{\psi_i}  \\
&&- \frac{N}{2} \log |\mathbf{\Phi}| - \frac{1}{2} {\rm  tr} \left \{ \sum_{n=1}^N \mathbf{\Phi}^{-1} E[\bm{F}_n\bm{F}_n^T|\bm{x}_n] \right\} - N   \sum_{i=1}^p\sum_{j=1}^m \rho P(|\lambda_{ij}|) 
\end{eqnarray*}
For given $\mathbf{\Lambda}_{\rm old}$, $\mathbf{\Psi}_{\rm old}$ and $\mathbf{\Phi}_{\rm old}$, the posterior  $f(\bm{f}_n | \bm{x}_n,\mathbf{\Lambda}_{\rm old}, \mathbf{\Psi}_{\rm old}, \mathbf{\Phi}_{\rm old})$ is normally distributed with $E[\bm{F}_n|\bm{x}_n] = \mathbf{M}^{-1}\mathbf{\Lambda}_{\rm old}^T\mathbf{\Psi}_{\rm old}^{-1} \bm{x}_n$ and $E[\bm{F}_n\bm{F}_n^T|\bm{x}_n] = \mathbf{M} ^{-1} + E[\bm{F}_n|\bm{x}_n] E[\bm{F}_n|\bm{x}_n] ^T$, where $\mathbf{M} = \mathbf{\Lambda}_{\rm old}^T\mathbf{\Psi}_{\rm old}^{-1}\mathbf{\Lambda}_{\rm old} + \mathbf{\Phi}^{-1}_{\rm old}$.  Then, we have
\begin{eqnarray*}
\sum_{n=1}^N E[\bm{F}_n]x_{ni} &=& \sum_{n=1}^N \mathbf{M}^{-1}\mathbf{\Lambda}_{\rm old}^T\mathbf{\Psi}_{\rm old}^{-1}\bm{x}_nx_{ni}=N\mathbf{M}^{-1}\mathbf{\Lambda}_{\rm old}^T\mathbf{\Psi}_{\rm old}^{-1}\mathbf{s}_i,\\
\sum_{n=1}^N E[\bm{F}_n\bm{F}_n^T] &=& \sum_{n=1}^N (\mathbf{M} ^{-1} + \mathbf{M}^{-1}\mathbf{\Lambda}_{\rm old}^T\mathbf{\Psi}_{\rm old}^{-1}\bm{x}_n\bm{x}_n^T\mathbf{\Psi}_{\rm old}^{-1}\mathbf{\Lambda}_{\rm old}\mathbf{M}^{-1})\\
&=&N (\mathbf{M} ^{-1} + \mathbf{M}^{-1}\mathbf{\Lambda}_{\rm old}^T\mathbf{\Psi}_{\rm old}^{-1}\mathbf{S}\mathbf{\Psi}_{\rm old}^{-1}\mathbf{\Lambda}_{\rm old}\mathbf{M}^{-1}),
\end{eqnarray*}
Let $\mathbf{M}^{-1}\mathbf{\Lambda}_{\rm old}^T\mathbf{\Psi}_{\rm old}^{-1}\mathbf{s}_i$ and $\mathbf{M} ^{-1} + \mathbf{M}^{-1}\mathbf{\Lambda}_{\rm old}^T\mathbf{\Psi}_{\rm old}^{-1}\mathbf{S}\mathbf{\Psi}_{\rm old}^{-1}\mathbf{\Lambda}_{\rm old}\mathbf{M}^{-1}$ be  $\mathbf{b}_i$ and $\mathbf{A}$, respectively.  Then, the expectation of  $\l_{\rho}^{C} $ in (\ref{ECL}) can be derived.

}
{
\baselineskip=6mm
\newpage
\bibliographystyle{ECA_jasa} 
\bibliography{paper-ref} 
\end{document}